\documentclass[review]{elsarticle}

\usepackage{hyperref}
\usepackage{mathtools}
\usepackage{siunitx}
\usepackage[percent]{overpic}
\usepackage{array}

\usepackage{url}
\usepackage{amsfonts}
\usepackage{booktabs}

\makeatletter
\def\ps@pprintTitle{%
   \let\@oddhead\@empty
   \let\@evenhead\@empty
   \def\@oddfoot{\reset@font\hfil\thepage\hfil}
   \let\@evenfoot\@oddfoot
}
\makeatother

\begin{document}

\begin{frontmatter}

\title{End-to-End Deep Residual Learning with Dilated Convolutions for Myocardial Infarction Detection and Localization}

\author{Iv\'an L\'opez-Espejo}
\address{Department of Electronic Systems, Aalborg University, Denmark\\
\texttt{ivl@es.aau.dk}}

\begin{abstract}
In this report, I investigate the use of end-to-end deep residual learning with dilated convolutions for myocardial infarction (MI) detection and localization from electrocardiogram (ECG) signals. Although deep residual learning has already been applied to MI detection and localization, I propose a more accurate system that distinguishes among a higher number (i.e., six) of MI locations. Inspired by speech waveform processing with neural networks, I found a more robust front-end than directly arranging the multi-lead ECG signal into an input matrix consisting of the use of a single one-dimensional convolutional layer per ECG lead to extract a \emph{pseudo}-time-frequency representation and create a compact and discriminative input feature volume. As a result, I end up with a system achieving an MI detection and localization accuracy of 99.99\% on the well-known Physikalisch-Technische Bundesanstalt (PTB) database.
\end{abstract}


\end{frontmatter}


\section{Introduction}
\label{sec:intro}

In this work I investigate the use of end-to-end deep residual learning with dilated convolutions for myocardial infarction (MI) detection and localization from electrocardiogram (ECG) signals.

Deep residual learning, originally conceived for image recognition applications, was proposed by He \emph{et al.} in \cite{He16} in order to deal with the performance degradation that occurs in convolutional neural networks (CNNs) when these are too deep. In short, the idea behind deep residual learning is as follows. Let $l$ designate a particular layer of a neural network in such a manner that $\mathbf{x}_{l-1}$ represents its input. The authors of \cite{He16} stated that it might be easier to optimize the residual mapping $\mathcal{H}_{l}^{l+k}(\mathbf{x}_{l-1})=\mathcal{F}_{l}^{l+k}(\mathbf{x}_{l-1})+\mathbf{x}_{l-1}$ between layers $l$ and $l+k$ ($k\in\mathbb{N}$) than the original mapping $\mathcal{F}_{l}^{l+k}(\mathbf{x}_{l-1})$ when networks are too deep. Such a residual mapping can be straightforwardly achieved by means of identity mapping, namely, by using identity shortcut connections skipping $k+1$ layers.

Deep residual learning has been applied in a successful manner to different tasks such as speaker verification \cite{Shi18}, keyword spotting \cite{Tang18,Lopez19} and, indeed, MI detection and localization \cite{Strodthoff19}. It is worth being aware of the following two major differences between \cite{Strodthoff19} and my proposal:
\begin{enumerate}
 \item Apart from healthy subjects (i.e., with no MI), the authors of \cite{Strodthoff19} only distinguish between two rough classes colloquially designated as anterior myocardial infarction (aMI) and inferior myocardial infarction (iMI). Instead, I design a more accurate system that distinguishes, apart from healthy subjects, among six different possible locations of the MI: anterior, antero-lateral, antero-septal,  inferior, infero-lateral and infero-postero-lateral. Notice that the first and last three locations above can be considered subclasses of aMI and iMI, respectively.
 \item In \cite{Strodthoff19}, the multi-lead ECG signal is arranged into a matrix that serves as input to the first two-dimensional convolutional layer of the deep residual learning architecture. Inspired by speech waveform processing with neural networks \cite{Sainath15}, I found a more robust front-end consisting of the use of one-dimensional convolutional layers to obtain an intermediate two-dimensional representation for each lead of the ECG signal. These intermediate representations are then stacked across the depth dimension and the resulting feature volume is fed to the first two-dimensional convolutional layer of my deep residual neural network, which, also unlike \cite{Strodthoff19}, uses dilated convolutions \cite{Yu16} and has a different structure.
\end{enumerate}
As a result of the above, I end up with a system the MI detection and localization accuracy of which is significantly higher than that of \cite{Strodthoff19}.

One-dimensional convolutional layers for the definition of feature extraction branches on a lead basis are also used for MI detection and localization in \cite{Liu18}. However, remarkable differences between \cite{Liu18} and my work can be found. I use a single one-dimensional convolutional layer per ECG lead to extract a \emph{pseudo}-time-frequency representation (see Section \ref{sec:system} for further details) and create a compact and discriminative input feature volume to exploit the strengths of a deep residual learning architecture. On the other hand, the authors of \cite{Liu18} ---which distinguish among five different possible locations of the MI, that is, one less than me--- use up to three one-dimensional convolutional and three more mean-pooling layers per ECG lead providing a vector to a much simpler classifier consisting of a fully-connected layer with softmax activation. The relatively high computational complexity of the feature extraction branches in \cite{Liu18} involves that the number of parameters of this system is above (according to my estimation from \cite{Liu18}) 6$\times$10$^{3}$, which is approximately the number of parameters of my deep residual learning-based proposal. In summary, I am able to achieve near flawless MI detection and localization performance with less computational complexity as well as by considering an additional MI location (i.e., infero-postero-lateral) with respect to \cite{Liu18}.

The rest of this report is organized as follows. In Section \ref{sec:system}, my end-to-end deep residual learning architecture with dilated convolutions for MI detection and localization is described. Then, implementation and training details are presented in Section \ref{sec:training}. Finally, in Section \ref{sec:results}, experimental results are shown.

\section{End-to-End System Description}
\label{sec:system}

\begin{figure}
\includegraphics[width=\linewidth]{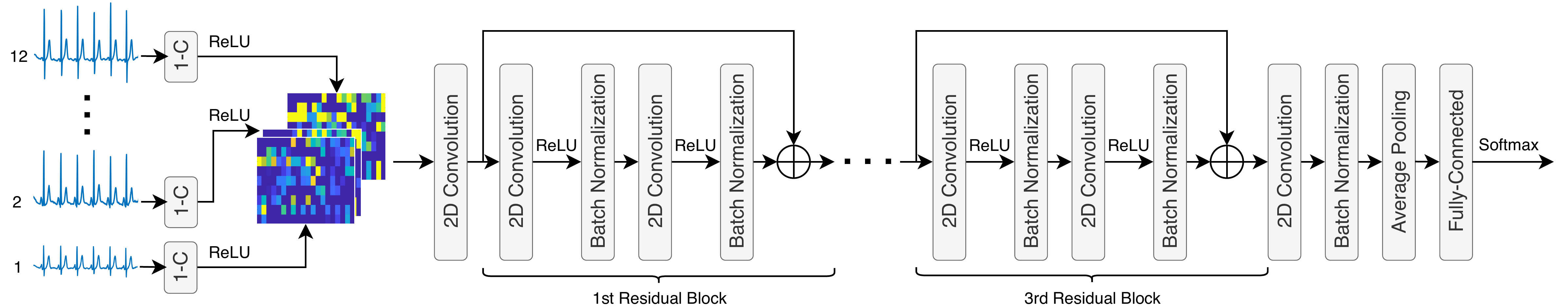}
\caption{Diagram of the proposed end-to-end deep residual neural network for myocardial infarction detection and localization. 1-C stands for one-dimensional convolutional layer.}
\label{fig:resnet}
\end{figure}

The proposed deep residual learning architecture for myocardial infarction (MI) detection and localization is based, although with significant differences, on those intended for keyword spotting presented in \cite{Tang18} and \cite{Lopez19}, which achieve impressive results.

A diagram of my end-to-end deep residual neural network with dilated convolutions can be seen in Figure \ref{fig:resnet}. As introduced in Section \ref{sec:intro}, inspired by speech waveform processing with neural networks \cite{Sainath15}, I consider a front-end based on one-dimensional convolutional layers, which, to the best of my knowledge, is novel in the context of MI detection and localization. According to my preliminary experiments, I found that this front-end is much more robust, and, therefore, provides better detection and localization performance, than directly arranging the multi-lead ECG signal into a matrix to be fed to the first two-dimensional convolutional layer of the deep residual neural network as in \cite{Strodthoff19}. Thus, in this work, 5 second long ECG signal segments are processed on a lead basis\footnote{In this report I work with 12-lead ECG signals.} by one-dimensional convolutional layers with 20 filters each, zero bias vectors, a kernel size of 100 samples and a stride of 50 samples. It is interesting to note that this set-up yields a \emph{pseudo}-time-frequency representation of the ECG signal, where the number of filters of the layers would correspond to the number of frequency bins and the kernel size and stride match the length of the analysis window and its hop size, respectively. Therefore, as the sampling rate of my ECG signals is 100 Hz, that is equivalent to using 20 \emph{pseudo}-frequency bins and a 1 second long analysis window with 50\% overlap\footnote{It is worth mentioning that, in accordance with my preliminary experiments, variations in the number of filters, the kernel size and the stride of the one-dimensional convolutional layers do not have a great impact on MI detection and localization performance.}. As a result, after the application of a rectified linear unit (ReLU) activation function, each lead is represented by a 9$\times$20 \emph{pseudo}-time-frequency matrix. Then, the twelve \emph{pseudo}-time-frequency matrices (one from each of the 12 leads) are stacked across the depth dimension and the resulting 9$\times$20$\times$12 feature volume is fed to the first two-dimensional convolutional layer of my architecture.

After the shallowest two-dimensional convolutional layer, there is a total of three residual blocks with identity mapping. Two convolutional layers, each of them followed by a ReLU activation function and a batch normalization layer for regularization purposes \cite{luo2018towards}, can be found in every residual block. The six convolutional layers of the residual blocks apply dilated convolutions the purpose of which is to increase the receptive field of the network \cite{Yu16}. The dilation rate of each of these layers depends on its position within the network. Thus, the dilation rates of the convolutional layers in the first and second residual blocks are $(1,1)$ and $(2,2)$, respectively. Additionally, the dilation rates of the first and second convolutional layers of the third residual block are, respectively, $(4,4)$ and $(8,8)$. A non-residual convolutional layer with $(16,16)$ convolution dilation, another batch normalization layer and an average pooling layer are appended to the third residual block. Then, for MI detection and classification, a fully-connected layer with softmax activation is employed. Finally, it is worth mentioning that all the two-dimensional convolutional layers have 7 filters each (equal to the number of neurons in the fully-connected layer), zero bias vectors and a kernel size of 3$\times$3.

\section{Implementation and Training Details}
\label{sec:training}

Keras \cite{chollet2015keras} was used to implement the end-to-end deep residual learning architecture depicted in Figure \ref{fig:resnet}. In practice, the parameters of the twelve one-dimensional convolutional layers are tied together during the learning process. The number of parameters of my system is only 5,997 in contrast to the deep residual learning architectures on which that is based, i.e., \cite{Tang18} and \cite{Lopez19}, which have more than 200,000 parameters.

After random initialization of parameters, deep residual learning models were trained for a total of 20 epochs ---which is more than enough for convergence--- by considering a multi-class cross-entropy loss function. I used the Adam optimizer \cite{Adam} with default parameters, that is, the learning rate and the learning rate decay were set to 0.001 and 0, respectively, $\beta_1=0.9$ and $\beta_2=0.999$. Moreover, the size of the minibatch was set to 32 training examples. For MI detection and localization, accuracy (i.e., the ratio between the number of correct predictions and the total number of predictions) was considered as performance metric.

\section{Experimental Results}
\label{sec:results}

\begin{figure}
\centering
\includegraphics[width=0.48\linewidth]{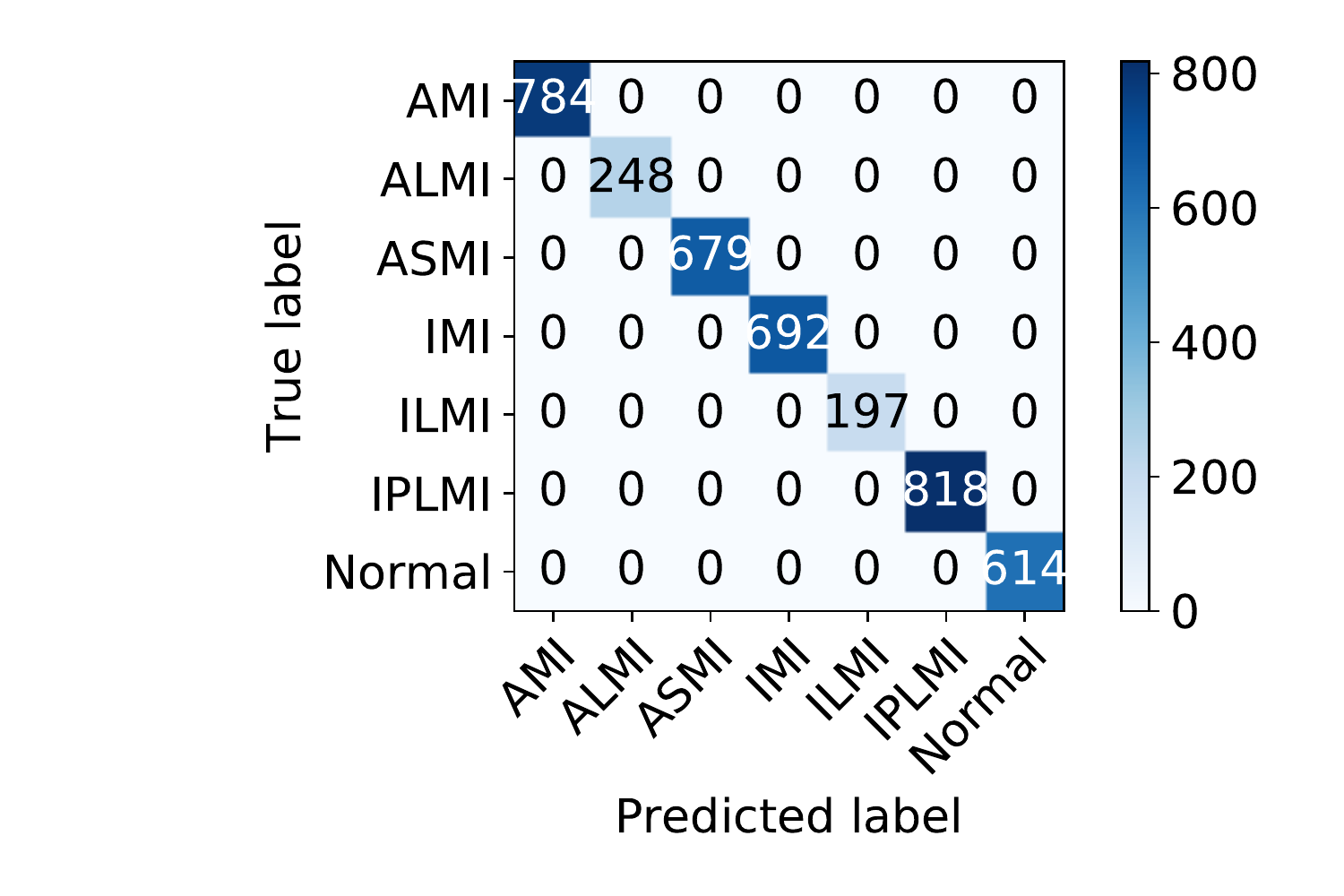} \includegraphics[width=0.48\linewidth]{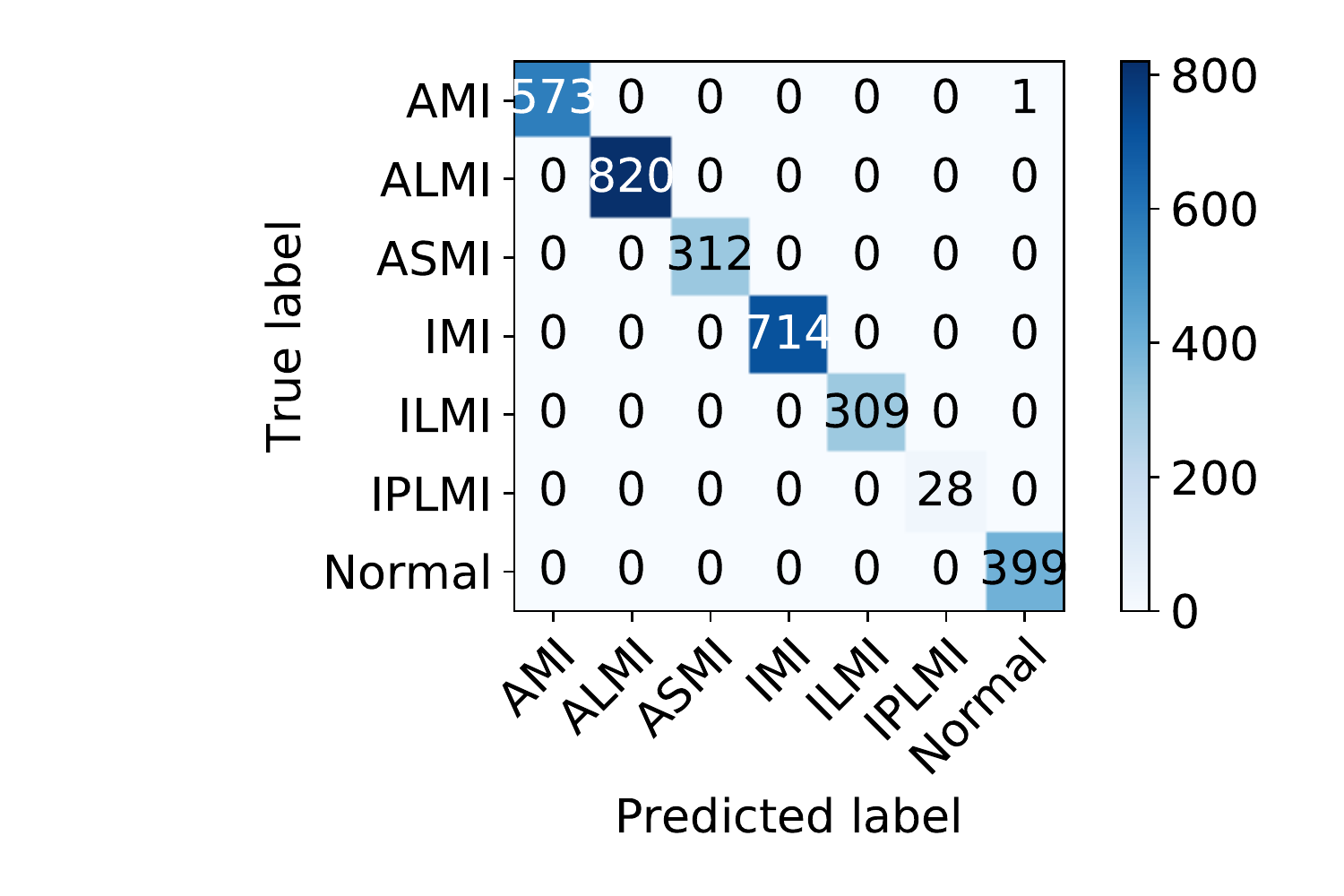}\\
\includegraphics[width=0.48\linewidth]{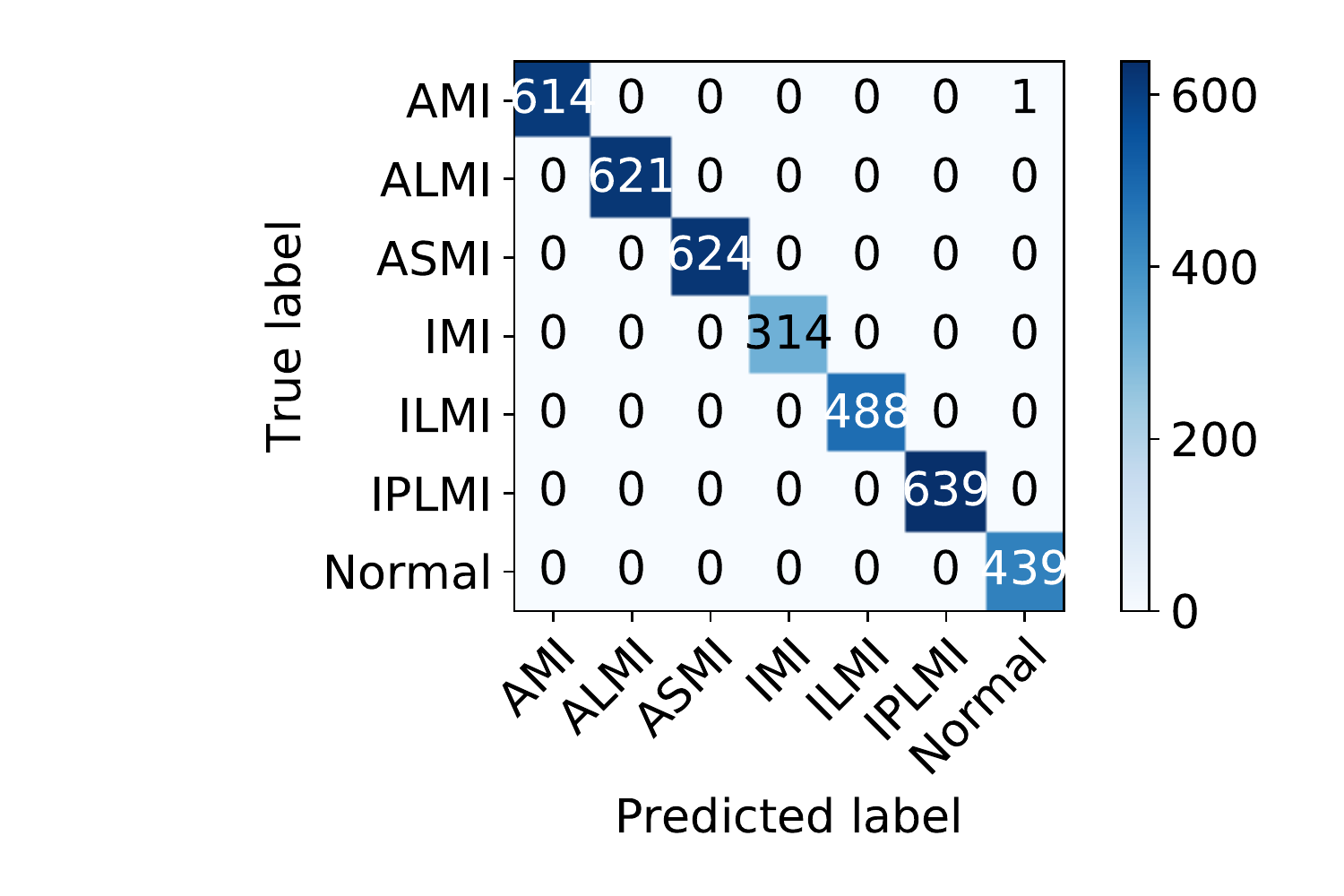} \includegraphics[width=0.48\linewidth]{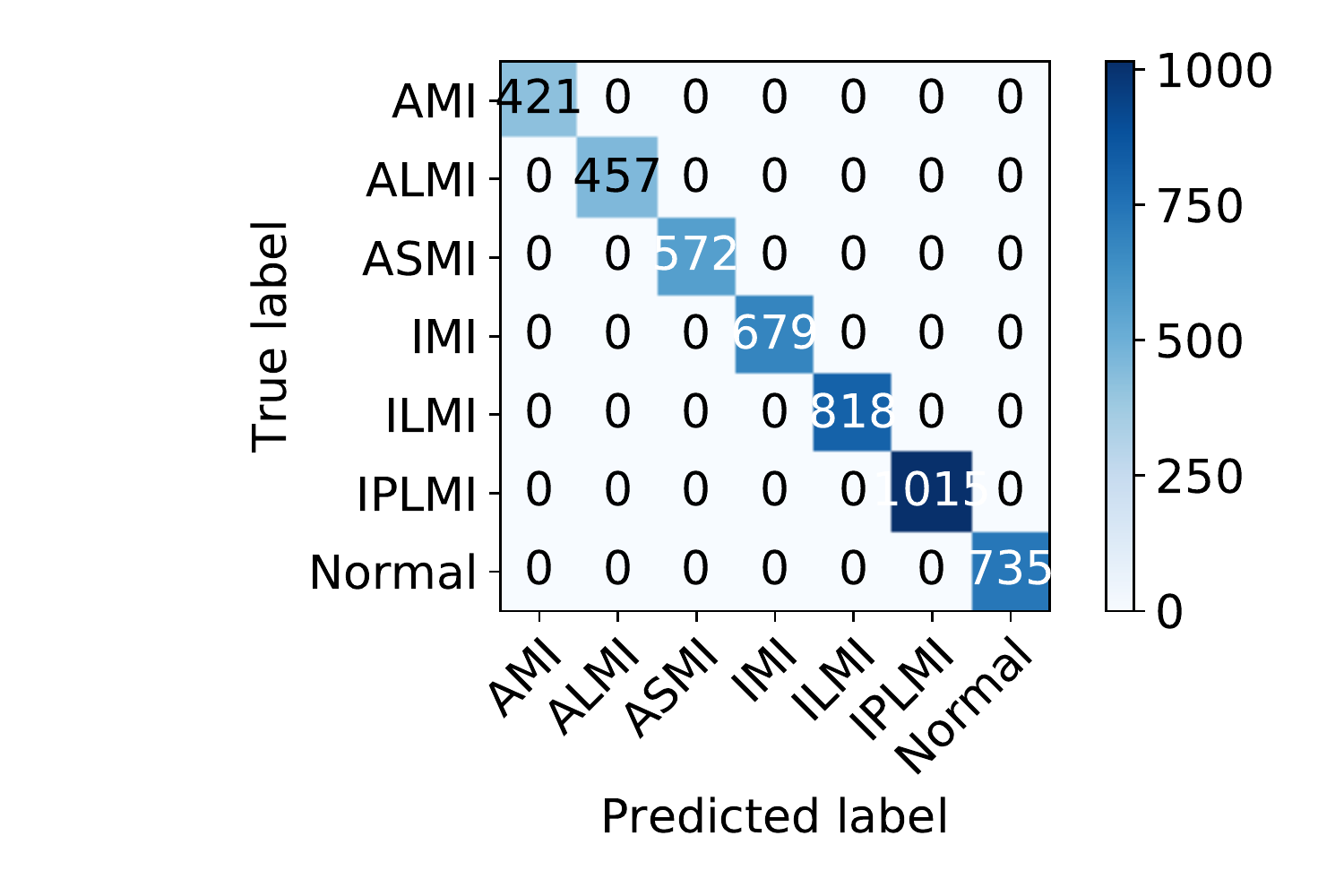}\\
\includegraphics[width=0.48\linewidth]{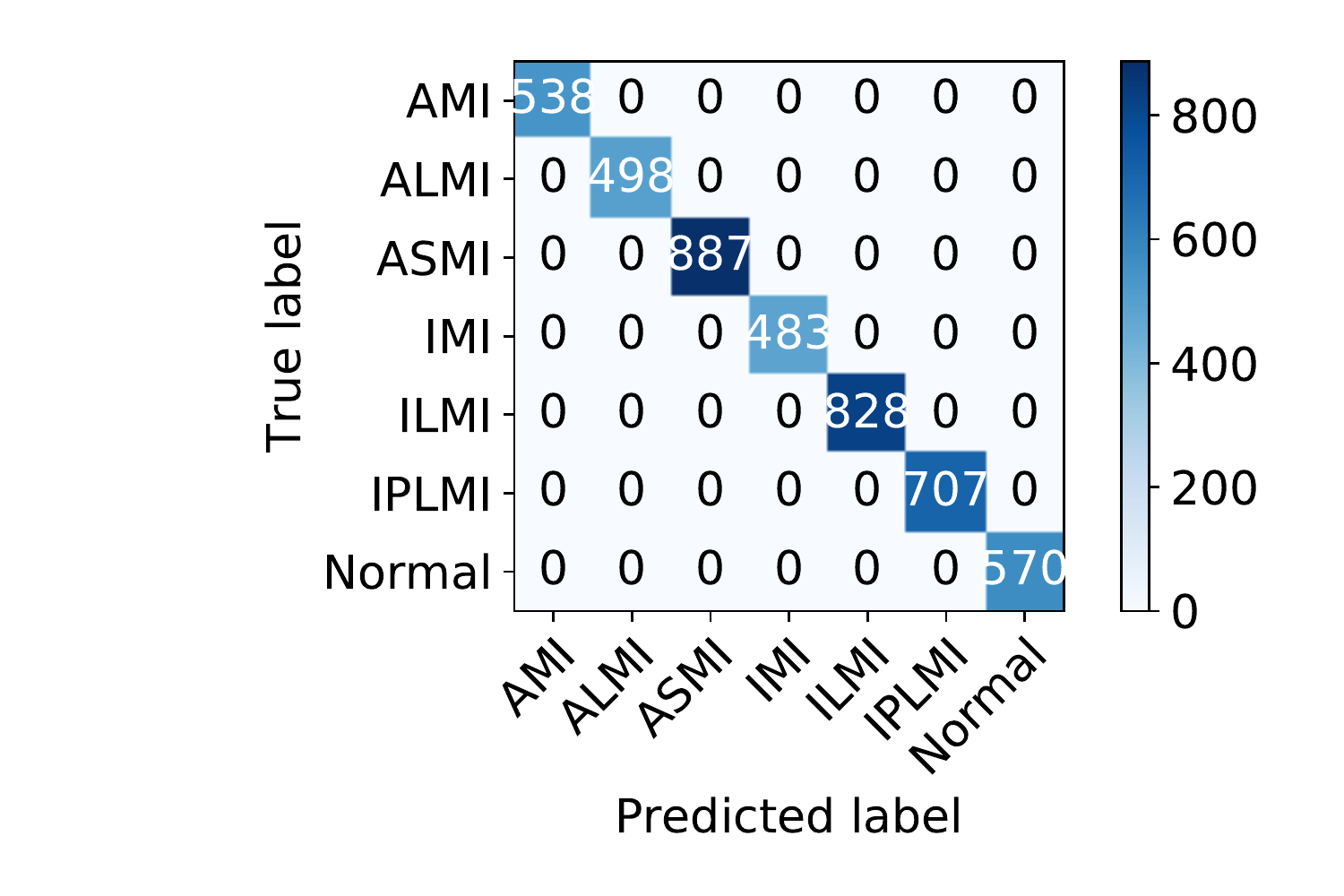}
\caption{Confusion matrices from myocardial infarction detection and localization on the PTB database. From left to right and top to bottom: confusion matrices from the 1st, 2nd, 3rd, 4th and 5th folds.}
\label{fig:confusion}
\end{figure}

For evaluation of the end-to-end system described in this report I use the Physikalisch-Technische Bundesanstalt (PTB) database \cite{PTB}. ECG data from 121 infarcted and 53 healthy subjects are employed. Both types of subjects are grouped and shuffled. Then, around 60\%, 10\% and 30\% of them are used to define training, validation and test sets in such a manner that, for a fair and more realistic evaluation, subjects do not overlap across sets. Accuracy results, in percentages, and confusion matrices are shown in Table \ref{tab:results} and Figure \ref{fig:confusion}, respectively, where I have considered 5-fold cross-validation. The average (i.e., across folds) myocardial infarction detection and localization accuracy with a 95\% confidence interval is 99.99\%$\pm$0.02.

\begin{table}
  \begin{center}
    \caption{Myocardial infarction detection and localization accuracy results, in percentages, on the PTB database when considering 5-fold cross-validation.}
    \label{tab:results}
    \resizebox{0.6\textwidth}{!}{\begin{tabular}{l|ccccc}
      \toprule
      \textbf{Fold} & 1 & 2 & 3 & 4 & 5 \\
      \midrule
      \textbf{Accuracy (\%)} & 100 & 99.97 & 99.97 & 100 & 100 \\
      \bottomrule
    \end{tabular}}
  \end{center}
\end{table}

\bibliography{mybibfile}

\begin{thebibliography}{10}
\expandafter\ifx\csname url\endcsname\relax
  \def\url#1{\texttt{#1}}\fi
\expandafter\ifx\csname urlprefix\endcsname\relax\def\urlprefix{URL }\fi
\expandafter\ifx\csname href\endcsname\relax
  \def\href#1#2{#2} \def\path#1{#1}\fi

\bibitem{He16}
K.~He, X.~Zhang, S.~Ren, J.~Sun, Deep residual learning for image recognition,
  in: Proceedings of {CVPR} 2016 -- Conference on Computer Vision and Pattern
  Recognition, June 26-July 1, Las Vegas, USA, 2016, pp. 770--778.

\bibitem{Shi18}
X.~Shi, M.~Zhu, X.~Du, End-to-end residual {CNN} with {L-GM} loss speaker
  verification system, in: Proceedings of {DSP} 2018 -- 23\textsuperscript{rd}
  IEEE International Conference on Digital Signal Processing, November 19-21,
  Shanghai, China, 2018.

\bibitem{Tang18}
R.~Tang, J.~Lin, Deep residual learning for small-footprint keyword spotting,
  in: Proceedings of {ICASSP} 2018 -- 43\textsuperscript{rd} IEEE International
  Conference on Acoustics, Speech and Signal Processing, April 15-20, Calgary,
  Canada, 2018, pp. 5484--5488.

\bibitem{Lopez19}
I.~L\'opez-Espejo, Z.-H. Tan, J.~Jensen, Keyword spotting for hearing assistive
  devices robust to external speakers, in: Proceedings of {INTERSPEECH} 2019 --
  20\textsuperscript{th} Annual Conference of the International Speech
  Communication Association, September 15-19, Graz, Austria, 2019.

\bibitem{Strodthoff19}
N.~Strodthoff, C.~Strodthoff, Detecting and interpreting myocardial infarction
  using fully convolutional neural networks, Physiological Measurement 40
  (2019).

\bibitem{Sainath15}
T.~N. Sainath, R.~J. Weiss, A.~Senior, K.~W. Wilson, O.~Vinyals, Learning the
  speech front-end with raw waveform {CLDNNs}, in: Proceedings of {INTERSPEECH}
  2015 -- 16\textsuperscript{th} Annual Conference of the International Speech
  Communication Association, September 6-10, Dresden, Germany, 2015.

\bibitem{Yu16}
F.~Yu, V.~Koltun, Multi-scale context aggregation by dilated convolutions, in:
  Proceedings of {ICLR} 2016 -- 4\textsuperscript{th} International Conference
  on Learning Representations, May 2-4, San Juan, Puerto Rico, 2016.

\bibitem{Liu18}
W.~Liu, Q.~Huang, S.~Chang, H.~Wang, J.~He, Multiple-feature-branch
  convolutional neural network for myocardial infarction diagnosis using
  electrocardiogram, Biomedical Signal Processing and Control 45 (2018) 22--32.

\bibitem{luo2018towards}
P.~Luo, X.~Wang, W.~Shao, Z.~Peng, Towards understanding regularization in
  batch normalization, in: Proceedings of {ICLR} 2019 -- 7\textsuperscript{th}
  International Conference on Learning Representations, May 6-9, New Orleans,
  USA, 2019.

\bibitem{chollet2015keras}
F.~Chollet, et~al., Keras, \url{https://keras.io} (2015).

\bibitem{Adam}
D.~P. Kingma, J.~Ba, Adam: A method for stochastic optimization, in:
  Proceedings of {ICLR} 2015 -- 3\textsuperscript{rd} International Conference
  on Learning Representations, May 7-9, San Diego, USA, 2015.

\bibitem{PTB}
R.~Bousseljot, D.~Kreiseler, A.~Schnabel, Nutzung der {EKG-S}ignaldatenbank
  {CARDIODAT} der {PTB} \"{u}ber das {I}nternet, Biomedizinische Technik 40
  (1995).

\end{thebibliography}

\end{document}